\documentclass[10pt,onecolumn,a4paper]{IEEEtran}
\usepackage{amsmath, amssymb}
\usepackage{graphicx}
\usepackage{multirow}
\usepackage{multicol}
\renewcommand{\baselinestretch}{1.5}

\newcommand{\figuramedia}[3]
{
\begin{figure}
  \centering
 \includegraphics[width=14cm]{#1}
  \caption{#2}\label{#3}
\end{figure}
}

\newcommand{\figuragrossa}[3]
{
\begin{figure}
  \centering
 \includegraphics[width=14cm]{#1}
  \caption{#2}\label{#3}
\end{figure}
}

\newcommand{\figura}[3]
{
\begin{figure}
  \centering
 \includegraphics[width=14cm]{#1}
  \caption{#2}\label{#3}
\end{figure}
}
\title{A Novel Proportional Fairness Criterion for Throughput
Allocation in Multirate IEEE 802.11}
\author{\authorblockN{M. Laddomada, F. Mesiti, M.
Mondin, and F. Daneshgaran
\thanks{Massimiliano Laddomada is with the Electrical Engineering Dept. of Texas A\&M University-Texarkana, email: \textrm{mladdomada@tamut.edu}.}
\thanks{F. Mesiti and M. Mondin are with DELEN, Politecnico di Torino,
Italy.}}
\thanks{F. Daneshgaran is with ECE Dept., California State University,
Los Angeles, USA.}}
\begin{document}
\maketitle
\begin{abstract}
This paper focuses on multirate IEEE 802.11 Wireless LAN employing
the mandatory Distributed Coordination Function (DCF) option. Its
aim is threefold. Upon starting from the multi-dimensional
Markovian state transition model proposed by Malone
\textit{et.al.} for characterizing the behavior of the IEEE 802.11
protocol at the Medium Access Control layer, it presents an
extension accounting for packet transmission failures due to
channel errors. Second, it establishes the conditions under which
a network constituted by $N$ stations, each station transmitting
with its own bit rate, $R^{(s)}_d$, and packet rate, $\lambda_s$,
can be assumed loaded. Finally, it proposes a modified
Proportional Fairness (PF) criterion, suitable for mitigating the
\textit{rate anomaly} problem of multirate loaded IEEE 802.11
Wireless LANs, employing the mandatory DCF option. Compared to the
widely adopted assumption of saturated network, the proposed
fairness criterion can be applied to general loaded networks.

The throughput allocation resulting from the proposed algorithm is
able to greatly increase the aggregate throughput of the DCF, while
ensuring fairness levels among the stations of the same order as the
ones guaranteed by the classical PF criterion.

Simulation results are presented for some sample scenarios,
confirming the effectiveness of the proposed criterion for
optimized throughput allocation.
\end{abstract}
%Fig.~\ref
%
\section{Introduction}
Consider the IEEE802.11 Medium Access Control (MAC)
layer~\cite{standard_DCF_MAC} employing the DCF based on the Carrier
Sense Multiple Access Collision Avoidance CSMA/CA access method. The
scenario envisaged in this work considers $N$ contending stations;
each station generates data packets with constant rate $\lambda_s$
by employing a bit rate, $R_d^{(s)}$, which depends on the channel
quality experienced. In this scenario, it is known that the DCF is
affected by the so-called \textit{performance anomaly}
problem~\cite{heusse}: in multirate networks the aggregate
throughput is strongly influenced by that of the slowest contending
station.

After the landmark work by Bianchi~\cite{Bianchi}, who provided an
analysis of the saturation throughput of the basic 802.11 protocol
assuming a two dimensional Markov model at the MAC layer, many
papers have addressed almost any facet of the behaviour of DCF in
a variety of traffic loads and channel transmission conditions.

Contributions proposed in the literature so far can be classified
in two main classes, namely \textit{DCF Modelling} and \textit{DCF
Throughput and Fairness Optimization}.\\

\noindent \textit{DCF modelling.} This is the topic that received
the most attention in the literature since the work by
Bianchi~\cite{Bianchi}.
    Papers \cite{HCheolLee}-\cite{Chatzimisios} model the influence of
real channel conditions on the throughput of the DCF operating in
saturated traffic conditions, while
\cite{zorzi_rao}-\cite{Spasenovski} thoroughly analyze the influence
of capture on the throughput of wireless transmission systems. Paper
\cite{Daneshgaran_sat} investigates the saturation throughput of
IEEE 802.11 in presence of non ideal transmission channel and
capture effects. The behavior of the DCF of IEEE 802.11 WLANs in
unsaturated traffic conditions has been analyzed in
\cite{Liaw}-\cite{daneshgaran_linmodel}. In \cite{Qiao}, the authors
look at the impact of channel induced errors and of the received
Signal-to-Noise Ratio (SNR) on the achievable throughput in a system
with rate adaptation, whereby the transmission rate of the terminal
is modified depending on either direct or indirect measurements of
the link quality.

Multirate modeling of the DCF has received some attention quite
recently \cite{Joshi}-\cite{ergen} as well. In \cite{Joshi} an
analytical framework for analyzing the link delay of multirate
networks is provided. In
\cite{cantieni}-\cite{daneshgaran_multirate}, authors provide DCF
models for finite load sources with multirate capabilities, while in
\cite{DuckYongYang}-\cite{ergen} a DCF model for networks with
multirate stations is provided and the saturation throughput is
derived. Remedies to performance anomalies are also discussed. In
both previous works, packet errors are only due to collisions
among the contending stations.\\

\noindent \textit{DCF throughput and fairness optimization.} This is
perhaps the issue most closely related to the problem dealt with in
this paper. The main reason for optimizing the throughput allocation
of the 802.11 DCF is the behaviour of the basic DCF in heterogeneous
conditions, with stations transmitting at multiple rates: the same
throughput is reserved to any contending station irrespective of its
bit rate, with the undesired consequence that lowest bit rate
stations occupy the channel for most time with respect to high rate
stations \cite{banchs}. Furthermore, the optimization of the
aggregate throughput when different stations contend for the channel
with different bit rates cannot be done without considering an
appropriate fairness approach; the reason is that the optimum
throughput would be achieved when only the highest rate stations
access the channel \cite{banchs}. In order to face this problem, a
variety of throughput optimization techniques, which account for
fairness issues, have been proposed in the literature. Paper
\cite{banchs} proposes a proportional fairness throughput allocation
criterion for multirate and saturated IEEE 802.11 DCF by focusing on
the 802.11e standard. In papers \cite{TinnirelloI}-\cite{joshi2} the
authors propose novel fairness criteria, which fall within the class
of the time-based fairness criterion. Time-based fairness guarantees
equal time-share of the channel occupancy irrespective to the
station bit rate.

Paper \cite{babu_1} investigates the fairness issue in 802.11
multirate networks by analyzing various time-based fairness
criteria. It demonstrates that with equal time-share of the
channel occupancy among multirate stations, the throughput
achieved by a reference station in a multirate scenario with $N$
contending stations is equal to the throughput that the same
reference station would achieve in a single rate scenario when
contending with other $N-1$ stations with its same rate.
Furthermore, the authors prove that the proportional fairness
criterion corresponds to fair channel time allocation in a
multirate scenario.

The effect of the contention window size on the performance of the
DCF have been also investigated in \cite{jiang}-\cite{Khalaj} in a
variety of different scenarios. Finally, papers
\cite{choudhury_errors}-\cite{choudhury} have been devoted to the
throughput optimization of the underlined DCF by optimizing a
number of key parameters of the DCF, such as the minimum
contention window size or the packet size.

A common hypothesis employed in the literature regards the
saturation assumption, which sometimes does not fit quite well to
real network traffic conditions. In real networks, traffic is mostly
non-saturated, different stations usually operate with different
loads, i.e., they have different packet rates, while the
transmitting bit rate can also differ among the contending stations.
Channel conditions are far from being ideal and often packet
transmission has to be rescheduled until the data is correctly
received. Due to Rayleigh and shadow fading conditions, a real
scenario presents stations transmitting at different bit rates,
because of multirate adaptation foreseen at the physical layer of
WLAN protocols such as IEEE 802.11b. In all these situations the
common hypothesis, widely employed in the literature, that all the
contending stations have the same probability of transmitting in a
randomly chosen time slot, does not hold anymore.

The aim of this paper is to investigate the behaviour of the DCF in
the most general scenario of a multirate network, when all the
previous effects act jointly, as well as to present a proportional
fairness criterion which accounts for general loading conditions as
exemplified by the packet rate $\lambda_s$ of the contending
stations. Contrary to the aforementioned works available in the
literature, we assume that the $s$-th station generates data packets
with its own size, $PL^{(s)}$, with its own constant rate
$\lambda_s$ by employing a bit rate, $R_d^{(s)}$, which depends on
the channel quality experienced, and it employs a minimum contention
window with size $W_0^{(s)}$. Moreover, each station is in a proper
load condition, which is independent from the loading conditions of
the other contending stations. Notice that these hypotheses make the
model proposed in this work quite different from the ones available
in the literature, where the saturated condition is mostly adopted.
One consequence of the proposed analysis is that unloaded,
heterogeneous networks do not need any throughput allocation among
stations. We propose a theoretical framework in order to identify
whether a tagged station is saturated, given the traffic conditions
of the remaining stations. As a starting point for the derivations
that follow, we consider the bi-dimensional Markov model proposed in
\cite{Malone}, and present the necessary modifications in order to
deal with multirate stations, non ideal transmission channel
conditions, and different packet sizes among the contending
stations.

The rest of the paper is organized as follows.
Section~\ref{sec:multirate_model} provides the necessary
modifications to the Markov model proposed in \cite{Malone}, while
the employed traffic model is discussed in
Section~\ref{subsection_traffic_model}. Section~\ref{sec:unsat}
proposes an analytical framework able to verify whether a network
of $N$ contending stations is loaded. The novel proportional
fairness criterion is presented in Section~\ref{sec:optimization},
while Section~\ref{SimulationResults_Section} presents simulation
results of some sample network scenarios. Finally,
Section~\ref{Conclusions_Section} draws the conclusions.
\section{The Network Scenario: Overview of The Markovian Model Characterizing the DCF}
\label{sec:multirate_model}
%
%%
%\figuragrossa{chain_multirate.eps}{Markov chain for the contention
%model of the generic $s$-th station in general traffic conditions,
%based on the 2-way handshaking technique, considering the effects
%of channel induced errors, unloaded traffic conditions, and
%post-backoff.}{fig.chain}
%%
In \cite{Malone}, the authors derived a bi-dimensional Markov model
for characterizing the behavior of the DCF in heterogeneous
networks, where each station has its own traffic, which could be
finite and characterized by the parameter $\lambda$, expressing the
packet arrival rate. In order to deal with non-saturated conditions,
the traffic model is described by an exponentially distributed
packet inter-arrival process. In this paper we consider a more
general network than \cite{Malone}. Indeed, in the investigated
network, each station employs a specific bit rate, $R_d^{(s)}$, a
specific transmission packet rate, $\lambda_s$, transmits packets
with size $PL^{(s)}$, and it employs a minimum contention window
with size $W_0^{(s)}$, which can differ from the one specified in
the IEEE 802.11 standard \cite{standard_DCF_MAC} (these
modifications are at the very basis of the proportional fairness
criterion proposed in Section~\ref{sec:optimization}). A finite
retry limit is considered in order to avoid infinite number of
retries when bad channel conditions inhibit the station from
successful transmission.

For the sake of greatly simplifying the evaluation of the expected
time slots required by the theoretical derivations that follow, we
consider $N_c \leq N$ classes of channel occupancy
durations\footnote{This assumption relies on the observation that in
actual networks some stations might transmit data frames presenting
the same channel occupancy. As an instance, a station STA1
transmitting a packet of size 128 bytes at 1 Mbps occupies the
channel for the same time of a station STA2 transmitting a packet of
size 256 bytes at 2 Mbps.}. First of all, given the payload lengths
and the data rates of the $N$ stations, the $N_c$ duration-classes
are arranged in order of decreasing durations identified by the
index $d \in \{1, \cdots, N_c\}$, whereby $d=1$ identifies the
slowest class. Notice that in our setup a station is denoted fast if
it has a short channel occupancy. Furthermore, each station is
identified by an index $s \in \{1, \cdots, N\}$, and it belongs to a
unique duration-class. In order to identify the class of a station
$s$, we define $N_c$ subsets $n(d)$, each of them containing the
indexes of the $L_d=|n(d)|$ stations within $n(d)$, with $L_d \leq
N, \forall d$ and $\sum_{d=1}^{N_c}L_d = N$. As an example, $n(3) =
\{1,5,8\}$ means that stations 1, 5, and 8 belong to the third
duration-class identified by $d=3$, and $L_d=3$.
\subsection{Bi-dimensional Contention Markov Model}
The modified bi-dimensional Markov model describing the contention
process of the $s$-th station\footnote{In order to keep the
notation concise, we omit the apex $s$ over the probabilities
involved in the model.} in the network is shown in
Fig.~\ref{fig.chain}.

Let us elaborate. We consider an overall number of $r$ different
backoff stages, starting from the zero-th stage. The maximum
Contention Window (CW) size is $W_{max} = 2^{m}W_0^{(s)}$, with $m
\leq r$, whereas the notation $W_i = \min(2^m
W_0^{(s)},2^iW_0^{(s)})$ is used to identify the $i$-th contention
window size ($W_0^{(s)}$ is the minimum contention window size of
the $s$-th station). Notice that after the $m$-th stage, the
contention window size is fixed to $W_{max}$ for the remaining
$(r-m)$ stages, after which the packet is dropped. An additional
backoff stage, identified by $(P,-)$, with the same window size of
the zero-th stage, is considered on top of the chain in order to
account for the post-backoff stage entered by the station after a
successful packet transmission, or packet drop
\cite{standard_DCF_MAC}. Moreover, the state labelled $(P,0)$ in
Fig.~\ref{fig.chain}, is used for emulating unloaded traffic
conditions.

After the post-backoff stage, a station starts a new transmission
because a new packet is available in the queue, provided that the
channel is sensed idle for DIFS seconds. On the other hand, a new
zero-th stage backoff is employed if the channel is sensed busy.
Notice that the post-backoff stage is entered only if the station
has no longer packets to transmit after a packet transmission;
otherwise, a zero-th stage is started. Moreover, if a new packet
arrives during a post-backoff stage, the station moves into the
zero-th stage, as depicted in Fig.~\ref{fig.chain}. Indeed,
backoff stages from $0$ to $r$ assume that the station's queue
contains at least a packet waiting for transmission.

A packet transmission is attempted only in the states labelled
$(i,0)$, $\forall i=0,\ldots,r$, as well as in the state $(P,0)$
only if there is a packet in the queue and the channel is sensed
idle for DIFS seconds. In case of collision, or due to the fact that
transmission is unsuccessful because of channel errors, the backoff
stage is incremented and the station moves in the state $(i+1,k)$,
where $k=0, \dots, W_{i+1}-1$, with uniform probability
$P_{eq}/W_{i+1}$, whereby $P_{eq}$, i.e., the probability of
equivalent failed transmission, is defined as
$P_{eq}=1-(1-P_{e})(1-P_{col})=P_{col}+P_e-P_e\cdot P_{col}$.
Probabilities $P_{col}$ and $P_e$ are, respectively, the collision
and the packet error probabilities related to the $s$-th station.

The transition probabilities for the generic $s$-th station's Markov
process in Fig.~\ref{fig.chain} could be separated as summarized in
what follows, depending on whether transitions start from standard
backoff states or from post-backoff states.

\noindent\textbf{Backoff state transitions}
\begin{equation}\label{EquBackoffStateTransit}\small
\begin{array}{lcll}
P_{i,k|i,k+1}  & = & 1,
  & k \in [0, W^{(s)}_i-2], \quad i \in [0,r]\\
P_{P,k|i,0}    & = & \frac{(1-P_{eq})(1-q)}{W_0^{(s)}},
  & k \in [0, W_0^{(s)}-1], \quad i \in [0,r-1]   \\
P_{0,k|i,0}    & = & \frac{(1-P_{eq})q}{W_0^{(s)}},
  & k \in [0, W_0^{(s)}-1], \quad i \in [0,r-1]   \\
P_{P,k|r,0}    & = & \frac{(1-q)}{W_0^{(s)}},
  & k \in [0, W_0^{(s)}-1],\\
P_{0,k|r,0}    & = & \frac{q}{W_0^{(s)}},
  & k \in [0, W_0^{(s)}-1].
\end{array}
\end{equation}
The meaning of the underlined probabilities is as follows. The
first equation in~(\ref{EquBackoffStateTransit}) states that, at
the beginning of each slot time, the backoff time is decremented.
The second (third) equation accounts for the fact that after a
successful transmission, the station goes in post-backoff because
of an empty (non empty) queue. In both equations, $q$ is used to
identify the probability that the queue contains at least a packet
waiting for transmission after a time slot, and it will be better
defined in Section \ref{subsection_traffic_model}, where the
employed traffic model is described.
The fourth equation deals with the situation in which the station
has reached the retry limit and, after a packet transmission, the
buffer of the station is empty. In this situation, the station
moves in the post-backoff stage with an empty queue. The last
equation accounts for a scenario similar to the previous one with
the difference that, after the packet transmission, the queue is
not empty.

\noindent\textbf{Post-backoff state transitions}
\begin{equation}\small
\begin{array}{lcll}
P_{P,k|P,k+1} & = & (1-q)
  & k \in [0, W_0^{(s)}-2]\\
P_{0,k|P,k+1} & = & q
  & k \in [0, W_0^{(s)}-2]\\
P_{P,0|P,0}   & = & (1-q)
  & \\
P_{P,k|P,0}   & = & \frac{q P_i (1-P_{eq})(1-q)}{W_0^{(s)}}
  & k \in [0, W_0^{(s)}-1] \\
P_{0,k|P,0}   & = & q\frac{(1-P_i)+q P_i (1-P_{eq}) }{W_0^{(s)}}
  & k \in [1,W_0^{(s)}-1]  \\
P_{1,k|P,0}   & = & \frac{q P_i P_{eq}}{W_1^{(s)}}
  & k \in [1,W^{(s)}_1-1]  \\
\end{array}
\end{equation}
The meaning of the underlined probabilities is as follows. The
first equation states that the station remains in the post-backoff
stage because the queue is empty, whereas the second equation
accounts for a transition in the zero-th backoff stage because a
new packet arrives at the end of a backoff slot. The third
equation models the situation in which there are no packets
waiting for transmission, and the station remains in the state
$(P,0)$ (idle state).

The fourth equation deals with the situation in which the station is
in the idle state $(P,0)$, and, at the end of a backoff slot, a new
packet arrives in the queue. In this scenario, the packet is
successfully transmitted, the queue is empty, and the station moves
in another post-backoff stage. The term $P_i$ identifies the
probability that the channel is idle, and it is defined as follows
with respect to the $s$-th tagged station:
%
%i.e., the probability that all the stations except the
%tagged $t$-th station are idle, is defined as:
%
\[
P^{(s)}_i=\prod_{j=1,j\ne s}^{N}\left(1-\tau_j\right)
\]
The fifth equation accounts for a scenario similar to the previous
one, except that the station queue is not empty after the immediate
transmission of a packet or a situation of busy channel. The last
equation models the scenario in which the station goes from the idle
state $(P,0)$ to the first backoff stage because of a failure of the
immediate transmission of the packet arrived in the head of the
queue.
\subsection{Throughput Evaluation}
Next line of pursuit consists in finding the probability $\tau_s$
that the $s$-th station starts a transmission in a randomly chosen
time slot. Due to the lengthy algebra involved in the derivations
needed for solving the bidimensional Markov chain, the relation
that defines $\tau_s$ has been derived in an additional document
available at \cite{Companion_document}, whereas for conciseness we
show the final formula in (\ref{eq:tau_s}) (shown at the bottom of
this page), along with the other key probabilities needed in this
paper.
\begin{figure*}
\line(1,0){515} %\textwidth
%\begin{flushleft}
%
\begin{equation}
\small
\begin{array}{lcl}
\tau_{s} & = & q^2 (1-P_{eq}^{r+1}) W_0 \cdot b_{P,0} \\
b_{P,0}  & = & \left\{ \frac{W_0(W_0+1)}{2}(X_B + X_P) -
\left(\frac{q W_0 - (1-q)[1-(1-q)^{W_0}]}{q^2}\right)X_P + q
W_0\frac{q P_{eq} K +
(1-q)}{[1-(1-q)^{W_0}](1-q)} \right\}^{-1}  \\
K & = & \frac{1}{2}\left[2W_0\frac{1-(2P_{eq})^{m-1}}{1-2P_{eq}} +
\frac{1-P_{eq}^{m-1}}{1-P_{eq}}\right] + (W_m+1)P_{eq}^{m-1}
\frac{1-P_{eq}^{r-m+1}}{1-P_{eq}}  \\
X_B & = & q \frac{ W_0q^2 + (1-Pi)[1-(1-q)^{W_0}](1-q)}{W_0[1-(1-q)^{W_0}](1-q)}  \\
X_P & = & \frac{q^2}{1-(1-q)^{W_0}}\\
\end{array}
\label{eq:tau_s}
\end{equation}
%\footnotesize For sake of clarity, we omit the apex $(s)$ from
%every terms of the above. Such terms refers to a generic station
%$(s)$.
%
\line(1,0){515}
\end{figure*}
Given $\tau_{s}$ in (\ref{eq:tau_s}), we can evaluate the
aggregate throughput $S$ as follows:
\begin{equation}\label{eq:throughput_aggr}
  S = \sum_{s=1}^{N} S_s= \sum_{s=1}^{N}\frac{1}{T_{av}}P_s^{(s)} \cdot (1 - P_e^{(s)}) \cdot PL^{(s)}
\end{equation}
whereby $T_{av}$ is the expected time per slot, $PL^{(s)}$ is the
packet size of the $s$-th station, and $P_s^{(s)}$ is the
probability of successful packet transmission of the $s$-th
station:
\begin{equation} \label{eq:pSucc_s}
P_s^{(s)}  =  \tau_{s}\cdot \prod_{\substack{j = 1 \\ j \neq
s}}^{N} (1 - \tau_{j})
\end{equation}
The evaluation of the aggregate throughput in
(\ref{eq:throughput_aggr}) requires the knowledge of the expected
time per slot, $T_{av}$. Its evaluation is the focus of the next
section.
\subsection{Evaluation of the Expected Time per Slot}
The expected time per slot, $T_{av}$, can be evaluated by weighting
the times spent by a station in a particular state with the
probability of being in that state. First of all, we observe that
there are four different kinds of time slots, with four different
average durations:
\begin{itemize}
\item the idle slot, in which no station is transmitting over the channel, with average duration
$T_{I}$;
\item the collision slot, in which more than one station is attempting to gain access to the channel, with average duration
$T_{C}$;
\item the slot due to erroneous transmissions because of imperfect channel conditions, with average duration
$T_{E}$;
\item the successful transmission slot, with average duration $T_{S}$.
\end{itemize}
The expected time per slot, $T_{av}$, can be evaluated by adding
the four expected slot durations:
\begin{equation} \label{eq:TAV}
T_{av} = T_I + T_C + T_S + T_E.
\end{equation}
We will now evaluate $T_I$, $T_C$, $T_S$, and $T_E$.

Upon identifying with $\sigma$ an idle slot duration, and defining
with $P_{TR}$ the probability that the channel is busy in a slot
because at least one station is transmitting:
\begin{equation}\label{eq:ptr}\small
  P_{TR} = 1 - \prod_{s = 1}^{N} (1-\tau_s)
\end{equation}
the average idle slot duration can be evaluated as follows:
\begin{equation}\small
T_I = (1-P_{TR}) \cdot \sigma
\end{equation}
The average slot duration of a successful transmission, $T_S$, can
be found upon averaging the probability $P_s^{(s)}$ that only the
$s$-th tagged station is successfully transmitting over the channel,
times the duration $T_S^{(s)}$ of a successful transmission from the
$s$-th station:
\begin{equation} \label{eq:TS}\small
T_S = \sum_{s=1}^{N} P_s^{(s)}\left(1-P_e^{(s)}\right) \cdot
T_S^{(s)}
\end{equation}
Notice that the term $(1-P_e^{(s)})$ accounts for the probability
of packet transmission without channel induced errors.

Analogously, the average duration of the slot due to erroneous
transmissions can be evaluated as follows:
\begin{equation} \label{eq:TE}\small
T_E = \sum_{s=1}^{N} P_s^{(s)} \cdot P_e^{(s)} \cdot T_E^{(s)}
\end{equation}
Let us focus on the evaluation of the expected collision slot,
$T_C$. There are $N_c$ different values of the collision
probability $P_{C}^{(d)}$, depending on the class of the tagged
station identified by $d$. We assume that in a collision of
duration $T_C^{(d)}$ (class-$d$ collisions), only the stations
belonging to the same class, or to higher classes (i.e., stations
whose channel occupancy is lower than the one of stations
belonging to the tagged station indexed by $d$) might be involved.

In order to identify the collision probability $P_{C}^{(d)}$, let
us first define the following three transmission probabilities
($P_{TR}^{C(d)},P_{TR}^{H(d)}$, $P_{TR}^{L(d)}$) under the
hypothesis that the tagged station belongs to the class $d$.
Probability $P_{TR}^{L(d)}$ represents the probability that at
least another station belonging to a lower class transmits, and it
can be evaluated as
\begin{equation}\label{eq:ptrSlow}\small
  P_{TR}^{L(d)} = 1 - \prod_{i = 1}^{d-1} \prod_{s \in n(i)} (1-\tau_s)
\end{equation}
Probability $P_{TR}^{H(d)}$ is the probability that at least one station belonging to a higher
class transmits, and it can be evaluated as
\begin{equation}\label{eq:ptrFast}\small
  P_{TR}^{H(d)} = 1 - \prod_{i = d+1}^{N_c} \prod_{s \in n(i)} (1-\tau_s)
\end{equation}
Probability $P_{TR}^{C(d)}$ represents the probability that at least a station in the same class
$d$ transmits:
\begin{equation}\label{eq:ptrClass}\small
  P_{TR}^{C(d)} = 1 - \prod_{s \in n(d)} (1-\tau_s)
\end{equation}
Therefore, the collision probability for a generic class $d$ takes
into account only collisions between at least one station of class
$d$ and at least one station within the same class (internal
collisions) or belonging to higher class (external collisions).
Hence, the total collision probability can be evaluated as:
\begin{equation}\label{eq:pColClass}\small
  P_{C}^{(d)} = P_{C}^{I(d)} + P_{C}^{E(d)}
\end{equation}
whereby
\begin{eqnarray}\label{eq:pColClass_E}\tiny
  P_{C}^{I(d)} &=& (1-P_{TR}^{H(d)}) \cdot (1-P_{TR}^{L(d)}) \cdot \\
  &&\cdot \left[P_{TR}^{C(d)} - \sum_{s \in n(d)}\tau_s\prod_{j \in n(d), j \neq\ s} (1-\tau_j)\right]\nonumber
\end{eqnarray}
represents the internal collisions between at least two stations
within the same class $d$, while the remaining are silent, and
\begin{equation}\label{eq:pColClass_H}\small
  P_{C}^{E(d)} = P_{TR}^{C(d)} \cdot P_{TR}^{H(d)} \cdot (1-P_{TR}^{L(d)})
\end{equation}
concerns to the external collisions with at least one station of
class higher than $d$.

Finally, the expected duration of a collision slot is:
\begin{equation} \label{eq:TC}\small
T_C = \sum_{d=1}^{N_c} P_C^{(d)} \cdot T_C^{(d)}
\end{equation}
Constant time durations $T_S^{(s)}$, $T_E^{(s)}$ and $T_C^{(d)}$
are defined in a manner similar to \cite{daneshgaran_multirate}
with the slight difference that the first two durations are
associated to a generic station $s$, while the latter is
associated to each duration class, which depends on the
combination of both payload length and data rate of the station of
class $d$.
\subsection{Traffic Model}
\label{subsection_traffic_model}
The employed traffic model for each station assumes a Poisson
distributed packet arrival process, whereby the inter-arrival times
among packets are exponentially distributed with mean $1/\lambda_t$.
In order to greatly simplify the analysis, we consider small queue,
as proposed in \cite{Malone}, even though the proposed analysis may
be easily extended to queues with any length. The traffic of each
station is accounted for within the Markov model by employing a
probability\footnote{A superscript $(t)$ is used for discerning the
probability $q$ among the stations.}, $q$, that accounts for the
scenario whereby at least one packet is available in the queue at
the end of a slot. In our setting, each station is characterized by
its own traffic, and the probability $q^{(t)}$ of the $t$-th station
can be evaluated by averaging over the four types of time slots,
namely idle, success, collision, and channel error time slot. Upon
noticing that, with the underlined packet model, the probability of
having at least one packet arrival during time $T$ is equal to $1 -
e^{-\lambda_t \cdot T}$, $q^{(t)}$ can be evaluated as:
\begin{equation}\label{eq:qs}
\begin{array}{rcl}
q^{(t)} & = & (1-P_{TR})\cdot (1 - e^{-\lambda_{t} \cdot \sigma}) + \\
&  & + \sum_{s=1}^{N} P_s^{(s)}\left(1-P_e^{(s)}\right) \cdot (1 - e^{-\lambda_{t} \cdot T_S^{(s)}}) + \\
&  & + \sum_{s=1}^{N} P_s^{(s)} \cdot P_e^{(s)} \cdot (1 - e^{-\lambda_{t} \cdot T_E^{(s)}}) + \\
&  & + \sum_{d=1}^{N_c} P_C^{(d)} \cdot (1 - e^{-\lambda_{t} \cdot
T_C^{(d)}})
\end{array}
\end{equation}
whereby the probabilities $P_{TR}$, $P_s^{(s)}$, and $P_C^{(d)}$
are, respectively, as defined in (\ref{eq:ptr}),
(\ref{eq:pSucc_s}), and (\ref{eq:pColClass}), whereas $P_e^{(s)}$
is the packet error rate of the $s$-th station.
\section{Evaluating the Network Loading Conditions}
\label{sec:unsat}
In a previous paper \cite{Daneshgaran_unsat}, we proved that the
behaviour of the aggregate throughput in a network of $N$
homogeneous\footnote{By homogeneous we simply mean that the
network is characterized by $N$ stations transmitting with the
same bit rate (no multirate hypothesis) and the same load.}
contending stations is a linear function of the packet arrival
rate $\lambda$ with a slope depending on both the number of
contending stations and the average payload length. We also
derived the interval of validity of the proposed model by showing
the presence of a critical $\lambda$, above which all the stations
begin operating in saturated traffic conditions.

This kind of behaviour, with appropriate generalizations, is also
observed when multirate and variable loaded stations are present in
the network. We have to identify a set of conditions for a network
to be considered as loaded. We notice in passing that this framework
is not generally considered in the literature, since most papers
assume saturated traffic conditions. A key observation from the
analysis developed in this section is that in an unloaded network
there is no need to guarantee fairness, since each contending
stations can transmit its packets at its own pace, regardless of its
minimum CW, as well as other network parameters.

Under the traffic model described in section
\ref{subsection_traffic_model}, we define unloaded a network in
which each contending station has a packet rate $\lambda_t$ less
than or equal to its packet service rate $\tilde{\mu}^{(t)}_S$:
\begin{equation}\label{saturation_condition}
\lambda_t\le \tilde{\mu}^{(t)}_S,~ \forall~ t\in \{1,\ldots,N\}
\end{equation}
The reason is simple: this condition ensures that the average packet
inter-arrival time is greater than or equal to the average service
time of the $t$-th station. In such a scenario, the probabilities of
collisions among stations are very low, and each contending station
is able, on the average, to gain the access to the channel as soon
as a new packet arrives in its queue. Notice that
$\tilde{\mu}^{(t)}_S$ only depends on the packet rates $\lambda_i$
of the other $N-1$ stations other than the tagged one.

The evaluation of the packet service rate $\tilde{\mu}^{(t)}_S$ in a
multirate and heterogeneous network is quite difficult \cite{Malone}
since packet arrivals may occur during the stage of post-backoff, as
well as during the usual backoff stages accomplished by each station
before gaining the channel for transmission. Since we are interested
in a threshold which differentiates the unsaturated from the
saturated loading conditions of the stations, we can employ an upper
bound defined by the \textit{saturation} service rate, identified as
$\mu^{(t)}_S$, in place of the actual service rate
$\tilde{\mu}^{(t)}_S$. The advantage relies on the observation that
such a bound is always evaluated considering a post-backoff stage.
Indeed, after a packet transmission, a new packet is always
available in the queue assuming saturated traffic; therefore, the
service time starts from a post-backoff phase whereby the contention
window is $W_0^{(s)}$. We notice that the saturation service time
always includes the post-backoff stage, thus its duration is longer
than the actual service time evaluated without considering the
post-backoff time.
%From this reasoning, it
%is possible to demonstrate that $\mu^{(t)}_S \leq
%\tilde{\mu}^{(t)}_S$.

Hence, in the remaining part of this section, we evaluate the
saturation service rate $\mu^{(t)}_S=1/T^{(t)}_{serv}$, i.e., the
rate at which packets are taken from the queue of the $t$-th station
under saturated conditions.

Upon considering the tagged station identified by the index $t \in
\{1,\cdots,N\}$, the saturation service time $T_{serv}^{(t)} $ can
be defined as follows \cite{babu_2}:
%
%%
%\figura{lambda_mu_s_caso1.eps}{Behaviour of the ratio
%$\lambda_t/\mu_S^{(t)}=\lambda_t\cdot T^{(t)}_{serv}$ in
%(\ref{saturation_condition}) in a network of three contending
%stations (labelled S1, S2 and S3) as a function of the packet rate
%$\lambda_1$ of the slowest station S1 transmitting at 1 Mbps. The
%other two stations transmit at 11 Mbps with constant packet rates,
%respectively equal to $100 \textrm{pkt/s}$ and $500 \textrm{pkt/s}$.
%The packet size PL is equal to 1028 bytes for the three contending
%stations.}{lambda_mu_s_caso1}
%%
%
\begin{equation}\label{tserv}\small
\begin{array}{rcl}
T_{serv}^{(t)}=\left\{\sum_{i=0}^{r} (P^{(t)}_{eq})^i \left( iT_C +
\sum_{j=0}^{i}\overline{W}^{(t)}_{j}\cdot T^{(t)}_{bo} + T^{(t)}_S
\right) + \right.&&
\\
\left. \underbrace{(P^{(t)}_{eq})^{r+1} \left( (r+1)T_C +
\sum_{j=0}^{r} \overline{W}^{(t)}_{j}
\cdot T^{(t)}_{bo}\right)}_{DROP}\right\} / \sum_{j=0}^{r+1} (P^{(t)}_{eq})^j&&\\
\end{array}
\end{equation}
The first term in the summation represents the average time that a
station spends through the backoff stages from $0$ to $r$ before
transmitting a packet, i.e., the so called MAC access time. We
notice that for the $i$-th stage, $i$ collisions of average
duration $T_C$, as well as $i$ backoff stages from $0$ to $i$
(each of them with an average number $\overline{W}^{(t)}_{j}$ of
slot of duration $T^{(t)}_{bo}$) occurred, after which the packet
is successfully transmitted with duration $T^{(t)}_s$. The second
term of the summation takes into account the average duration of a
packet drop that occurs after $(r+1)$ collisions and backoff
stages. The whole summation is scaled by a normalization factor
that takes into account the probability set over which the service
time is evaluated.

The average number of slots for the $i$ backoff stages, is defined
as
%
%%
%\figura{throughput_caso1.eps}{Behaviour of the per-station
%throughput in a network of three contending stations (labelled S1,
%S2 and S3) as a function of the packet rate $\lambda_1$ of the
%slowest station S1 transmitting at 1 Mbps. The other two stations
%transmit at 11 Mbps with constant packet rates, respectively equal
%to $100 \textrm{pkt/s}$ and $500 \textrm{pkt/s}$. The packet size
%PL is equal to 1028 bytes for the three contending
%stations.}{throughput_caso1}
%%
$$\overline{W}^{(t)}_{j}=(2^{\min(j,m)}\cdot W^{(t)}_0-1)
/ 2.$$
Each slot has average duration $T^{(t)}_{bo}$, which is
substantially evaluated as $T_{av}$ in (\ref{eq:TAV}) except that
the tagged station $(t)$ is not considered because it is either
idle, or in a backoff state.

Let us discuss two sample scenarios in order to derive a variety of
observations that are at the very basis of the fairness problem
developed in the next section. The network parameters used in the
investigated IEEE802.11b MAC layer are reported in
Table~\ref{tab.design.times}~\cite{standard_DCF_MAC}. The first
investigated scenario considers a network of $3$ contending
stations. Two stations, namely S2 and S3, transmit packets with
constant rates $\lambda=100~\textrm{ pkt/s}$ and
$\lambda=500~\textrm{ pkt/s}$, respectively. The bit rate of the two
stations S2 and S3 is $11$ Mbps. The third station, S1, has a bit
rate equal to 1 Mbps and a packet rate $\lambda_1$ that is varied in
the range $[0,2000]$ pkt/s in order to investigate its effects on
the network load. The behaviour of the three ratios
$\lambda_t/\mu_S^{(t)}$ is shown in Fig.~\ref{lambda_mu_s_caso1}.
Some observations are in order. First of all, notice that as far as
S1 increases its packet rate, the curves $\lambda_2\cdot
T^{(2)}_{serv}$ and $\lambda_3\cdot T^{(3)}_{serv}$ tend to increase
because of the increasing values of the service times
$T^{(2)}_{serv}$ and $T^{(3)}_{serv}$ experienced by S2 and S3.
Notice that, as $\lambda_1$ increases, the slowest station S1 tends
to transmit more often. The station S3 goes in loaded condition when
$\lambda_1\approx 20$pkt/s, while S2 can be considered loaded for
$\lambda_1\approx 200$ pkt/s. When a station becomes loaded, the
incoming packets tend to be stored in the station queue waiting for
transmission since the service rate, i.e., the number of packets
that on average are serviced by the MAC, is below the rate by which
the packets arrive in the station queue.

The per-station throughput achieved by the three stations in the
investigated scenario is shown in Fig.~\ref{throughput_caso1}. The
throughput gained by the two fastest stations, S2 and S3, tends to
decrease because of the anomaly problem in the multirate scenario:
the slowest station tends to occupy the channel longer and longer as
far as its packet rate $\lambda_1$ increases. In the same figure, we
show two tick curves. The horizontal line L2 corresponds to the
saturation throughput of S1, while the straight line L1 is the
tangent to the throughput curve passing through the origin. For very
small values of $\lambda_1$, the throughput of the station S1 grows
linearly with $\lambda_1$. Packets are mainly transmitted as soon as
they arrive at the MAC layer, and the station throughput is
approximately equal to $\lambda_1\cdot PL^{(1)}$. However, when the
station approaches the transition point $\lambda_1^*=89.19$ pkt/s
derived with the proposed framework (this is the value of
$\lambda_1$ corresponding to the relation
$\lambda_1/\mu^{(1)}_S=1$), the throughput curve tends to reach the
asymptote L2, which corresponds to the saturation throughput of S1.
Notice that the curve L2 approximately corresponds to 0.73 Mbps,
which is $\lambda_1^*\cdot PL^{(1)}=89.19\cdot 1028\cdot 8$ bps.

In the second scenario, the stations S1 and S2 are interested by a
constant packet rate noticed in the label of
Fig.~\ref{lambda_mu_s_caso3}. These two stations do not get loaded
by the increasing packet rate of the station S3 since the curves
$\lambda_1/\mu^{(1)}_S$ and $\lambda_2/\mu^{(2)}_S$ are strictly
less than one. As a consequence, the per-station throughput of both
S1 and S2 is approximately constant across the range of values of
the packet rate $\lambda_3$ as noticed in
Fig.~\ref{throughput_caso3}. On the other hand, the throughput
achieved by the station S3 tends to saturate as soon as $\lambda_3$
reaches the value $533$ pkt/s noticed in
Fig.~\ref{lambda_mu_s_caso3}.
%
%"in an similar way, we can obtain the threshold for the station
%S2, which is 382.97 pkt/s.

In the light of the previous two sample scenario, let us summarize
the main ingredients of the results proposed in this section. As
observed in the two previous sample scenarios, this method allows to
identify whether the network is loaded by establishing the
thresholds of each contending station in the network. This issue has
been overlooked in the literature, where the saturation assumption
is widely adopted. Moreover, this issue is at the very basis of any
throughput optimization strategy since an unloaded network does not
need to be optimized.

We say that the network is loaded when every station has a traffic
above its proper threshold. On the other hand, it should be
noticed that a network can be unloaded even if a subset of the
stations is loaded. This was the case of the second scenario
described above, where, despite the fact that the station S3 was
interested by an increasing traffic load $\lambda_3$, the stations
S1 and S2 did not experience any performance loss (see
Fig.~\ref{throughput_caso3}) because their traffics were below the
respective thresholds.
%
%A well explaining example is
%represented in Fig.5 where S3 increase its traffic while the
%per-station throughput of S1 and S2 is almost not lowered.
%
%*** because their respective traffic (L1 = 10 pkt/s and L2 = 100 pkt/s) are well
%below their thresholds (L1* = 89.19 pkt/s and L2* = 382.97 pkt/s) *** in realta'
%non e' proprio corretto perche' se S3 aumenta il suo carico bisognerebbe ricalcolare le soglie.
%
%
%%
%\figura{lambda_mu_s_caso3.eps}{Behaviour of the ratio
%$\lambda_t/\mu_S^{(t)}=\lambda_t\cdot T^{(t)}_{serv}$ in
%(\ref{saturation_condition}) in a network of three contending
%stations (labelled S1, S2 and S3) as a function of the packet rate
%$\lambda_3$ of the fastest station S3 transmitting at 11 Mbps. The
%station S1 transmits with constant packet rate $10 \textrm{pkt/s}$
%at 1 Mbps, whereas the station S2 transmits with constant packet
%rate $100 \textrm{pkt/s}$ at 11 Mbps. The packet size PL is equal to
%1028 bytes for the three contending stations.}{lambda_mu_s_caso3}
%%
%%
%\figura{throughput_caso3.eps}{Behaviour of the per-station
%throughput in a network of three contending stations (labelled S1,
%S2 and S3) as a function of the packet rate $\lambda_3$ of the
%fastest station S3 transmitting at 11 Mbps. The other two stations
%transmit with constant packet rates, respectively equal to $10
%\textrm{pkt/s}$ and $100 \textrm{pkt/s}$. The packet size PL is
%equal to 1028 bytes for the three contending
%stations.}{throughput_caso3}
%%
%%%
%%
\section{The Proportional Fairness Throughput Allocation Algorithm}
\label{sec:optimization}
This section presents the novel throughput allocation criterion,
which aims at improving fairness among the $N$ contending stations.
In order to face the fairness problem in the most general scenario,
i.e., multirate DCF and general station loading conditions, we
propose a novel Proportional Fairness Criterion (PFC) by starting
from the PCF defined by Kelly in \cite{kelly97}, and employed in
\cite{banchs} in connection to proportional fairness throughput
allocation in multirate and saturated DCF operations.

It is known that one of the main drawbacks of the basic DCF
operating in a multirate scenario relies on the fact that it behaves
in such a way as to guarantee equal long-term channel access
probability to the various contending stations \cite{banchs,babu_1}.
In order to solve this problem, various optimization algorithms have
been proposed in the literature (see, for instance,
\cite{banchs}-\cite{QXia}). These contributions allowed to highlight
the behaviour of the DCF as well as various drawbacks when operating
in a multirate scenario. For instance, it is known that the
aggregate throughput of multirate IEEE802.11-like networks is
optimized when only the high rate stations transmit, while the low
rate stations are kept silent. Of course, this result is not
desirable from a fairness point of view, even though it mitigates
the DCF performance anomaly noticed in \cite{heusse}.

To the best of our knowledge, the solutions proposed so far in the
literature refer to homogeneous networks in that all the
contending stations operate with the same traffic, as exemplified
by the station packet rate $\lambda_s$, and channel conditions.
Furthermore, almost all the works focus on saturated traffic
conditions. As already mentioned before, in a practical setting
the contending stations have their own traffic and are affected by
different channel conditions. The key observation here is that a
fair throughput allocation should account for the station packet
rate, as well as for the specific channel conditions experienced
by each contending station.

Let us briefly mention the rationales at the very basis of the
PFC. A proportional fairness optimization criterion allocates to
each station a throughput \textit{proportional} to the station
transmission rate. Resorting to the notation proposed in
\cite{kelly97}, a throughput allocation vector
$\underline{x}=\{x_s; s=1,\cdots,N\}$ is \textit{proportional
fair} if the following condition holds:
\begin{equation} \label{eq:prop_fair}
\sum_{s=1}^{N} \frac{y_s^* - x_s}{x_s} \leq 0
\end{equation}
for any other feasible throughput allocation vector
$\underline{y}^*$. The PFC maximization problem, which satisfies
(\ref{eq:prop_fair}), can be formalized as follows:
\begin{eqnarray} \label{eq:max_prop_fair}
\mbox{max} && \sum_{s=1}^{N} \log(x_s)  \\
\mbox{over}&&  x_s \in [0,x_{s,m}],~ s=1,\ldots,N\nonumber
\end{eqnarray}
whereby $x_{s,m}$ is the maximum throughput of the $s$-th station.
Due to the strict concavity of the logarithmic function and
because of the compactness of the feasible region $x_s \in
[0,x_{s,m}],~ s=1,\ldots,N$, there exists a unique solution to the
optimization problem (\ref{eq:max_prop_fair}). This implies that a
local maximum is also global.

Given these preliminaries, in the proposed model, the traffic of
each station is characterized by the packet arrival rate
$\lambda_s$, which depends mainly on the application layer. Let
$\lambda_{max}$ be the maximum packet rate among
$\lambda_1,\ldots,\lambda_N$.

Consider the following modified optimization problem:
\begin{equation} \label{eq:max_prop_fair_lambda}
\begin{array}{ll}
\mbox{max} & U=U(S_1,\cdots,S_N)=\sum_{s=1}^{N} \frac{\lambda_s}{\lambda_{max}} \cdot \log(S_s)\\
\mbox{subject to}&  S_s \in [0,S_{s,m}],~ s=1,\ldots,N
\end{array}
\end{equation}
whereby $S_s$ is the throughput of the $s$-th station, and $S_{s,m}$
is its maximum value, which equals the station bit rate $R_d^{(s)}$.
In our scenario, the individual throughputs, $S_s$, are interlaced
because of the interdependence of the probabilities involved in the
transmission probabilities $\tau_s, \forall s=1,\ldots,N$. For this
reason, we reformulate the maximization problem in order to find the
$N$ optimal values of $\tau_s$ for which the cost function $U$ in
(\ref{eq:max_prop_fair_lambda}) gets maximized. The optimal values
$\tau_s^*$ are then used to set the network parameters of each
station.

Due to the compactness of the feasible region $S_s \in
[0,S_{s,m}],\forall s$, the maximum of $U(S_1,\cdots,S_N)$ can be
found among the solutions of $\nabla U=\left(\frac{\partial
U}{\partial \tau_1},\cdots,\frac{\partial U}{\partial
\tau_N}\right)=0.$ After some algebra (the derivations are
reported in the Appendix), the solutions can be written as:
\begin{equation} \label{eq:max_eq_tauj}
\frac{\lambda_j}{\lambda_{max}}
\frac{1}{\tau_j}-\frac{1}{1-\tau_j}\sum_{k=1, k\ne
j}^{N}\frac{\lambda_k}{\lambda_{max}}=\frac{C}{T_{av}}
\frac{\partial T_{av}}{\partial \tau_j} , \quad \forall
j=1,\ldots,N
\end{equation}
whereby $C=\sum_{i=1}^N \frac{\lambda_i}{\lambda_{max}}$, and
$T_{av}$ is a function of $\tau_1,\cdots,\tau_N$ as noticed in
(\ref{eq:TAV}).

Due to the presence of $T_{av}$, a closed form of the maximum of
$U(S_1,\cdots,S_N)$ cannot be found. Notice that it is quite
difficult to derive the contribution of the partial derivative of
$T_{av}$ on $\tau_j$, especially when $N\gg 1$, because of the huge
number of network parameters belonging to different stations. The
definition of $T_{av}$ in (\ref{eq:TAV}) is composed by four
different terms, which include the whole set of $\tau_s$, $\forall
s$. In order to overcome this problem, we first numerically obtain
the optimal values $\tau_s^*, \forall s$ from
(\ref{eq:max_eq_tauj}), then, we choose the value of the minimum
contention window size, $W^{(s)}_0$, by equating the optimizing
$\tau_s^*$ to (\ref{eq:tau_s}) for any $s$.

The optimization criterion summarized in
(\ref{eq:max_prop_fair_lambda}) will be denoted as Load
Proportional Fairness (LPF) criterion in the following.

Let us derive some observations on the proposed throughput
allocation algorithm by contrasting it to the classical PF
algorithm. Consider two contending stations with packet rates
$\lambda_1=50~ \textrm{pkt/s}$ and $\lambda_2=100~
\textrm{pkt/s}$, respectively. Employing the classical PF method,
a throughput allocation is proportionally fair if a reduction of
$x\%$ of the throughput allocated to one station is
counterbalanced by an increase of more than $x\%$ of the
throughputs allocated to the other contending stations.

In our setup, the ratio $\lambda_1/\lambda_2$ can be interpreted
as the frequency by which the first station tries to get access to
the channel relative to the other station. Therefore, a throughput
allocation is proportionally fair if, for instance, a reduction of
$20\%$ of the throughput allocated to the first station, which has
a relative frequency of $1/2$, is counterbalanced by an increase
of more than $40\%$ of the throughput allocated to the second
station. In a scenario with multiple contending stations, the
relative frequency is evaluated with respect to the station with
the highest packet rate in the network, that gets unitary relative
frequency.

Based on extensive analysis, we found that the optimization
problem (\ref{eq:max_prop_fair_lambda}) sometimes yields
throughput allocations that cannot be actually managed by the
stations. As a reference example, assume that, due to the specific
channel conditions experienced, the first station has a bit rate
equal to $1$ Mbps and needs to transmits $200~ \textrm{pkt/s}$.
Given a packet size of $1024$ bytes, that is $8192$ bits, the
first station would need to transmit $8192\times 200~
\textrm{bps}\approx 1.64 \textrm{Mbps}$ far above the maximum bit
rate decided at the physical layer. In this scenario, such a
station could not send over the channel a throughput greater than
1Mbps. The same applies to the other contending stations in the
network experiencing similar conditions. In order to face this
issue, we considered the following optimization problem
\begin{equation} \label{eq:max_prop_fair_lambda_trunc}
\begin{array}{ll}
\mbox{max} & \sum_{s=1}^{N} \frac{\lambda^*_s}{\lambda^*_{max}} \cdot \log(S_s)\\
\mbox{over}&  S_s \in [0,S_{s,m}],~ s=1,\ldots,N
\end{array}
\end{equation}
where $\forall s=1,\ldots,N$,
\[
\lambda^*_s=\left\{
\begin{array}{ll}
\lambda_s, & \textrm{if} ~\lambda_s\cdot PL^{(s)}\cdot 8 \le R_d^{(s)}\\
 \frac{R_d^{(s)}}{8\cdot PL^{(s)}},           & \textrm{if} ~\lambda_s\cdot PL^{(s)}\cdot 8 > R_d^{(s)}
\end{array} \right.
\]
and $\lambda^*_{max}=\max_{s}\lambda^*_s$. The allocation problem
in (\ref{eq:max_prop_fair_lambda_trunc}), solved as for the LPF in
(\ref{eq:max_prop_fair_lambda}), guarantees a throughput
allocation which is proportional to the frequency of channel
access of each station relative to their actual ability in
managing such traffic. The idea relies on the observation that it
is not optimal to allocate a throughput that the station cannot
actually manage.

The optimization criterion summarized in
(\ref{eq:max_prop_fair_lambda_trunc}) will be denoted as Modified
Load Proportional Fairness (MLPF) criterion in what follows. As in
the case of the LPF criterion, the MLPF optimization problem in
(\ref{eq:max_prop_fair_lambda_trunc}) is solved by first
numerically obtaining the optimal values $\tau_s^*, \forall s$,
and then choosing the value of the minimum contention window sizes
$W^{(s)}_0$ by equating the optimizing $\tau_s^*$ to
(\ref{eq:tau_s}) independently for any $s$.

\section{Simulation Results}
\label{SimulationResults_Section}
%%
%\begin{table}\caption{Typical network parameters}
%\begin{center}
%\begin{tabular}{c|c||c|c}\hline
%\hline MAC header & 28 bytes&Propag. delay $\tau_p$ & 1 $\mu s$\\
%\hline PLCP Preamble & 144 bit & PLCP Header & 48 bit \\
%\hline PHY header & 24 bytes&Slot time & 20 $\mu s$\\
%\hline PLCP rate & 1Mbps & W$_0$ & 32\\
%\hline No. back-off stages, m & 5 & W$_{max}$ & 1024\\
%\hline Payload size & 1028 bytes&SIFS & 10 $\mu s$\\
%\hline ACK & 14 bytes&DIFS & 50 $\mu s$\\
%\hline ACK timeout & 364$\mu s$ &EIFS & 364 $\mu s$\\
%
%\hline\hline
%\end{tabular}
% \label{tab.design.times}
%\end{center}
%\end{table}
%%
%%
%%
%\figura{behav_Tserv.eps}{Behaviour of the critical packet rates
%($\mu_S^{(t)}=1/T^t_{serv}$) in a network of three contending
%stations as a function of the packet rate $\lambda_1$ of the slowest
%station transmitting at 1 Mbps. The other two stations transmit with
%a constant packet rate equal to $500 \textrm{pkt/s}$ at 11 Mbps.
%Notice that curves $\mu_S^{(t)}$ related to the stations at 11 Mbps
%are superimposed since they both employ the same network
%parameters.}{behav_Tserv}
%%
%
This section presents simulation results obtained for a variety of
network scenarios optimized with the fairness criteria proposed in
the previous section.

We have developed a C++ simulator modelling both the DCF protocol
details in 802.11b and the backoff procedures of a specific number
of independent transmitting stations. The simulator considers an
Infrastructure BSS (Basic Service Set) with an Access Point (AP) and
a certain number of fix stations which communicate only with the AP.
Traffic is generated following the exponential distribution for the
packet interarrival times. Moreover, the MAC layer is managed by a
state machine which follows the main directives specified in the
standard~\cite{standard_DCF_MAC}, namely waiting times (DIFS, SIFS,
EIFS), post-backoff, backoff, basic and RTS/CTS access modes. The
typical MAC layer parameters for IEEE802.11b reported in
Table~\ref{tab.design.times}~\cite{standard_DCF_MAC} have been used
for performance validation.

The first investigated scenario, namely scenario A, considers a
network with $3$ contending stations. Two stations transmit packets
with rate $\lambda=500~\textrm{ pkt/s}$ at $11$ Mbps. The payload
size, assumed to be common to all the stations, is $PL=1028$ bytes.
The third station has a bit rate equal to 1Mbps and a packet rate
$\lambda=1000~\textrm{ pkt/s}$.

From (\ref{saturation_condition}), it is straightforward to notice
that the scenario A refers to a loaded network.
Fig.~\ref{behav_Tserv} shows the service rates $\mu_S^{(t)}$ of the
three contending stations as a function of the packet rate
$\lambda_1$ of the station transmitting at 1 Mbps. The operating
point of the considered scenario A is highlighted in
Fig.~\ref{behav_Tserv}. Notice that, since the service rates
$\mu_S^{(t)}$ of the three stations are below the respective packet
rates $\lambda_t$, the network is loaded. Moreover, the service
rates of the three stations tend to the same values because of the
rate anomaly problem: the station transmitting at 1 Mbps reduces the
service rates of the other stations.

The simulated normalized throughput achieved by each station in this
scenario is depicted in the left subplot of Fig.~\ref{fig_PF_1} for
the following four setups. The three bars labelled 1-DCF represent
the normalized throughput achieved by the three stations with a
classical DCF. The second set of bars, labelled 2-PF, identifies the
simulated normalized throughput achieved by the DCF optimized with
the PF criterion \cite{banchs,kelly97}, whereby the actual packet
rates of the stations are not considered. The third set of bars,
labelled 3-LPF, represents the normalized throughput achieved by the
three stations when the allocation problem
(\ref{eq:max_prop_fair_lambda}) is employed. Finally, the last set
of bars, labelled 4-MLPF, represents the simulated normalized
throughput achieved by the contending stations when the CW sizes are
optimized with the modified fairness criterion in
(\ref{eq:max_prop_fair_lambda_trunc}). Notice that the throughput
allocations guaranteed by LPF and MLPF improve over the classical
DCF. When the station packet rate is considered in the optimization
framework, a higher throughput is allocated to the first station
presenting the maximum value of $\lambda$ among the considered
stations. However, the highest aggregate throughput is achieved when
the allocation is accomplished with the optimization framework
4-MLPF. The reason for this behaviour is related to the fact that
the first station requires a traffic equal to $8.22\textrm{
Mbps}=10^3 \textrm{ pkt/s}\cdot 1028 \textrm{ bytes/pkt}\cdot 8
\textrm{ bits/pkt}$, which is far above the maximum traffic
($1\textrm{ Mbps}$) that the station would be able to deal with in
the best scenario. In this respect, the MLPF criterion results in
better throughput allocations since it accounts for the real traffic
that the contending station would be able to deal with in the
specific scenario at hand.
%%
%\figuramedia{fig_composita_1.eps}{Simulated normalized throughput
%achieved by three contending stations upon employing 1) a classical
%DCF; 2) DCF with PF allocation; 3) DCF optimized as noted in
%(\ref{eq:max_prop_fair_lambda}); and 4) DCF optimized with the MLPF
%criterion. Left and right plots refer to scenarios A and B,
%respectively.}{fig_PF_1}
%%

Similar considerations can be drawn from the results shown in the
right subplot of Fig.~\ref{fig_PF_1} (related to scenario B),
whereby in the simulated scenario the two fastest stations are
also characterized by a packet rate greater than the one of the
slowest station. Notice that the optimization framework 3-LPF is
able to guarantee improved aggregate throughput with respect to
both the non-optimized DCF and the classical PF algorithms. The
operating point of the scenario B is highlighted in
Fig.~\ref{behav_Tserv}; based on the considerations above, this is
a loaded network as well.

The aggregate throughputs achieved in the two investigated
scenarios are as follows:\\

\begin{tabular}{l|l||c|c|c|c}\hline
\hline
\multicolumn{2}{c||}{Scenarios} & 1-DCF & 2-PF & 3-LPF & 4-MLPF \\
\hline
\multirow{2}{*}{A} & Jain's Index & 0.451 & 0.872 & 0.724 & 0.848 \\
                   & S [Mbps] & 1.85 & 3.60 & 3.10 & 5.07 \\
\hline
\multirow{2}{*}{B} & Jain's Index & 0.474 & 0.881 & 0.976 & 0.847 \\
                   & S [Mbps] & 1.99 & 3.63 & 4.62 & 5.07\\

\hline\hline
\end{tabular}
\\\\

\noindent whereby we also show the fairness Jain's index \cite{jain}
evaluated on the normalized throughputs noted in the subplots of
Fig.~\ref{fig_PF_1}. It is worth noticing that the proposed MLPF
throughput allocation criterion is able to guarantee improved
aggregate throughput relative to both the classical DCF and the PF
algorithm, with fairness levels on the same order of the ones
guaranteed by the classical PF algorithm.
%
%%
%\figura{fig_composita_2.eps}{Simulated normalized throughput
%achieved by three contending stations as a function of the packet
%rate of the slowest station in DCF, LPF and MLPF
%modes.}{ThroughDCF_simfin}

For the sake of investigating the behaviour of the proposed
allocation criteria as a function of the packet rate of the slowest
station, we simulated the throughput allocated to a network composed
by three stations, whereby the slowest station, transmitting at
1Mbps, presents an increasing packet rate in the range 10$-$3000
pkt/s. The other two stations transmit packets at the constant rate
$\lambda=500$ pkt/s at 11 Mbps. The simulated throughput of the
three contending stations is shown in the three subplot of
Fig.~\ref{ThroughDCF_simfin} for the unoptimized DCF, as well as for
the two criteria LPF and MLPF. Some considerations are in order. Let
us focus on the throughput of the DCF (uppermost subplot in
Fig.~\ref{ThroughDCF_simfin}). As far as the packet rate of the
slowest station increases, the throughput allocated to the fastest
stations decreases quite fast because of the performance anomaly of
the DCF \cite{heusse}. The three stations reach the same throughput
when the slowest station presents a packet rate equal to
$500~\textrm{pkt/s}$, corresponding to the one of the other two
stations. From $\lambda=500~\textrm{pkt/s}$ all the way up to 3000
pkt/s, the throughput of the three stations do not change anymore,
since all the stations have a throughput imposed by the slowest
station in the network. Let us focus on the results shown in the
other two subplots of Fig.~\ref{ThroughDCF_simfin}, labelled LPF and
MLPF, respectively. A quick comparison among these three subplots in
Fig.~\ref{ThroughDCF_simfin} reveals that the MLPF allocation
criterion guarantees improved aggregate throughput for a wide range
of packet rates of the slowest station, greatly mitigating the rate
anomaly problem of the classical DCF operating in a multirate
setting. In terms of aggregate throughput, the best solution is
achieved with the MLPF criterion, which avoids that the slowest
station receives a throughput allocation that would not be able to
employ due to its reduced bit rate (1 Mbps).
%
%
%%
%\begin{table}\caption{Jain's Fairness Index and Aggregate Throughput S}
%\small
%\begin{center}
%\begin{tabular}{l|l||c|c|c|c}\hline
%\hline
%\multicolumn{2}{c||}{Scenarios in Fig.~\ref{fig_PF_1}} & 1-DCF & 2-PF & 3-LPF & 4-MLPF \\
%\hline
%\multirow{2}{*}{A} & Jain's Index & 0.451 & 0.872 & 0.724 & 0.848 \\
%                   & S [Mbps] & 1.85 & 3.60 & 3.10 & 5.07 \\
%\hline
%\multirow{2}{*}{B} & Jain's Index & 0.474 & 0.881 & 0.976 & 0.847 \\
%                   & S [Mbps] & 1.99 & 3.63 & 4.62 & 5.07\\
%
%\hline\hline
%\end{tabular}
% \label{JainsFairnessIndex}
%\end{center}
%\end{table}
%%
%
\section{Conclusions}
Focusing on multirate IEEE 802.11 Wireless LAN employing the
mandatory Distributed Coordination Function (DCF) option, this
paper established the conditions under which a network constituted
by a certain number of stations transmitting with their own bit
rates and packet rates can be considered loaded. It then proposed
a modified proportional fairness criterion suitable for mitigating
the \textit{rate anomaly} problem of multirate loaded IEEE 802.11
Wireless LANs.

Simulation results were presented for some sample scenarios
showing that the proposed throughput allocation was able to
greatly increase the aggregate throughput of the DCF, while
ensuring fairness levels among the stations of the same order of
the ones available with the classical proportional fairness
criterion.
\label{Conclusions_Section}
\appendix
\section{Maximum of the Aggregate Throughput}
\label{append_I}
The objective of this section is to derive the maximum of the
function $U(S_1,\cdots,S_N)$ in (\ref{eq:max_prop_fair_lambda}) as
a function of $\tau_s, \forall s=1,\ldots,N$.

Upon substituting (\ref{eq:pSucc_s}) and (\ref{eq:TAV}) in
(\ref{eq:throughput_aggr}), and deriving with respect to $\tau_j$,
we obtain:
\begin{eqnarray} \label{eq:max_deriv_j}
 \frac{\partial}{\partial \tau_j}  \sum_{i=1}^{N}
\frac{\lambda_i}{\lambda_{max}} \cdot
%
%& &
%\nonumber\\
%\scriptstyle
%
\log \left[ \frac{1}{T_{av}} \tau_i \prod_{\substack{k=1\\k \ne
i}}^N (1-\tau_k) (1-P_e^{(i)})PL^{(i)} \right] &=&
\nonumber \\
%
%\scriptstyle
%
\frac{\partial}{\partial \tau_j} \sum_{i=1}^{N}
\frac{\lambda_i}{\lambda_{max}} \cdot
%
%\nonumber\\
%\scriptstyle
%
\left[ \log \tau_i + \sum_{\substack{k=1\\k \ne i}}^N
\log(1-\tau_k) +\log(1-P_e^{(i)})+\right.&&\nonumber\\
\left. +\log(PL^{(i)}) - \log(T_{av}) \right] &=&
\nonumber\\
%
%\scriptstyle
%
\frac{\partial}{\partial \tau_j} \sum_{i=1}^{N}
\frac{\lambda_i}{\lambda_{max}} \cdot
%\nonumber\\
%\scriptstyle
%
\left[ \log \tau_i + \sum_{\substack{k=1\\k \ne
i}}^N\log(1-\tau_k)-\log(T_{av})\right]&&
\end{eqnarray}
whereby the last relation stems from the independence of
$P_e^{(i)}$ and $PL^{(i)}$ on $\tau_j$. Exchanging the derivative
with the summation yields:
\begin{equation} \label{eq:max_deriv_j_2}
\frac{\lambda_j}{\lambda_{max}}
\frac{1}{\tau_j}-\frac{1}{1-\tau_j}\sum_{k=1, k\ne
j}^{N}\frac{\lambda_k}{\lambda_{max}}-\frac{C}{T_{av}}
\frac{\partial T_{av}}{\partial \tau_j} , \quad \forall
j=1,\ldots,N
%
%\frac{\lambda_j}{\lambda_{max}}
%\left(\frac{1}{\tau_j}-\frac{N-1}{1-\tau_j}\right) - C \cdot
%\frac{\partial \log(T_{AV})}{\partial \tau_j}
\end{equation}
whereby $C=\sum_{i=1}^N \frac{\lambda_i}{\lambda_{max}}$. By
equating (\ref{eq:max_deriv_j_2}) to zero, (\ref{eq:max_eq_tauj})
easily follows.
\clearpage
\figuragrossa{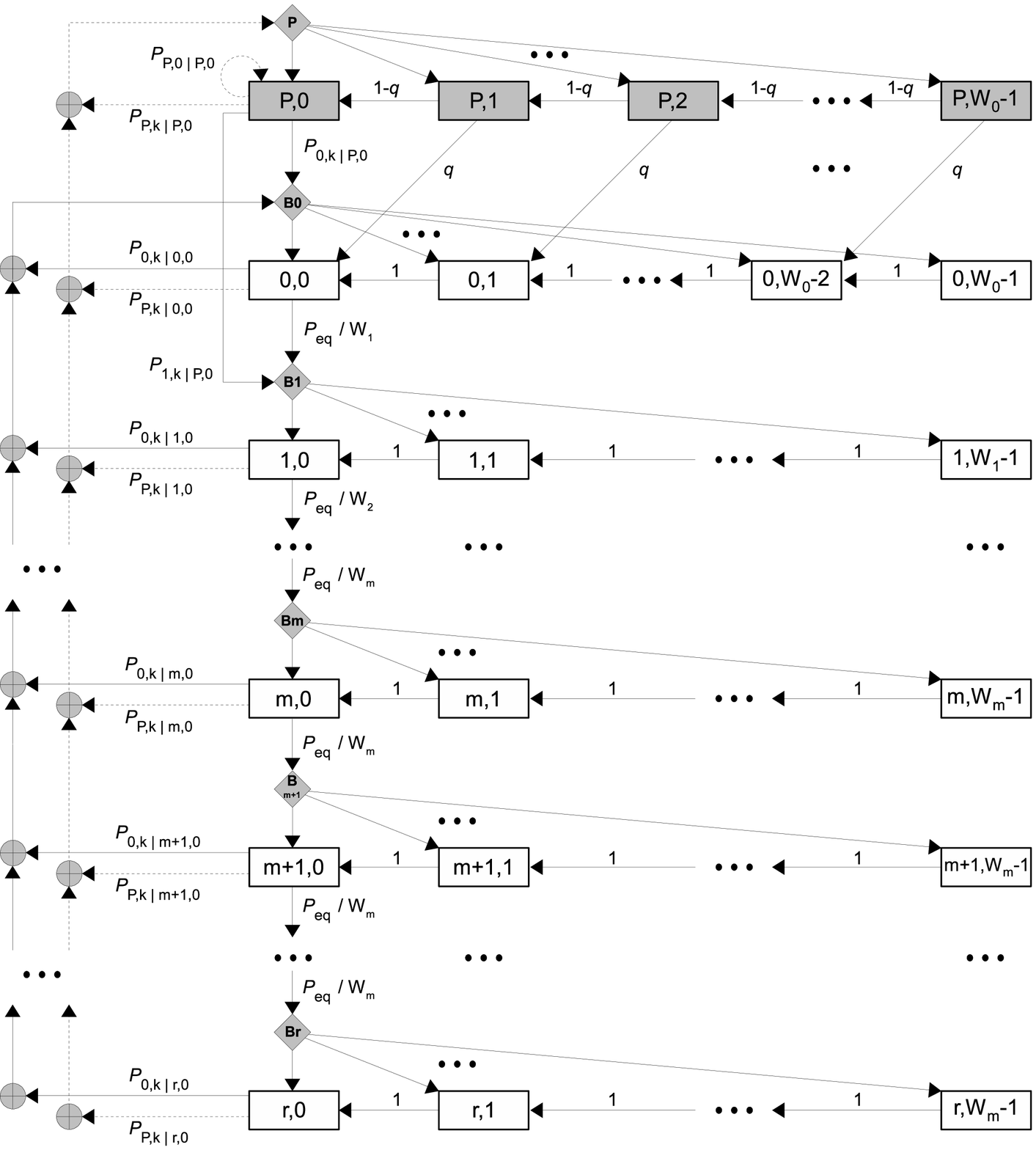}{Markov chain for the contention
model of the generic $s$-th station in general traffic conditions,
based on the 2-way handshaking technique, considering the effects
of channel induced errors, unloaded traffic conditions, and
post-backoff.}{fig.chain}
\clearpage
\figura{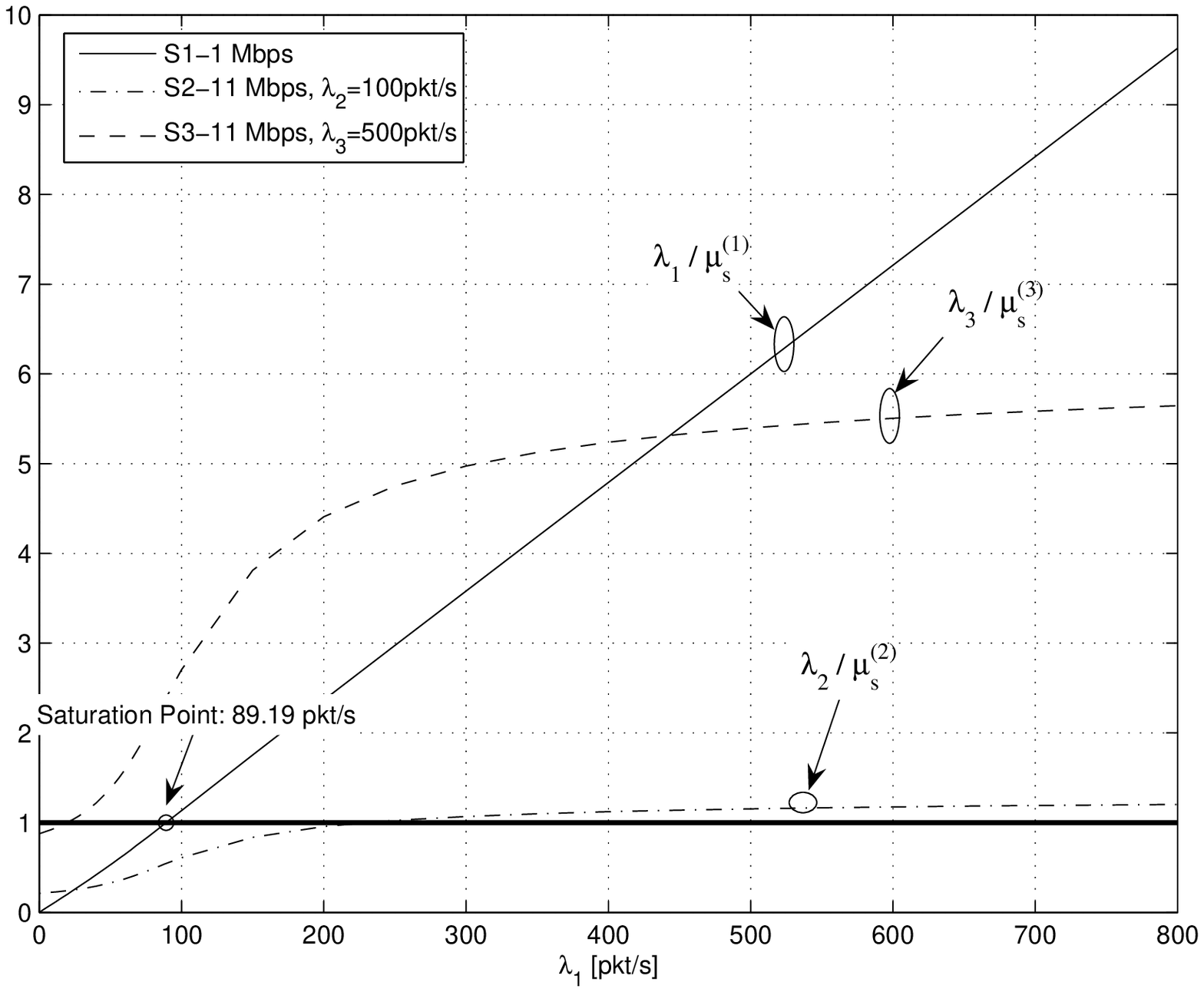}{Behaviour of the ratio
$\lambda_t/\mu_S^{(t)}=\lambda_t\cdot T^{(t)}_{serv}$ in
(\ref{saturation_condition}) in a network of three contending
stations (labelled S1, S2 and S3) as a function of the packet rate
$\lambda_1$ of the slowest station S1 transmitting at 1 Mbps. The
other two stations transmit at 11 Mbps with constant packet rates,
respectively equal to $100 \textrm{pkt/s}$ and $500
\textrm{pkt/s}$. The packet size PL is equal to 1028 bytes for the
three contending stations.}{lambda_mu_s_caso1}
\clearpage
\figura{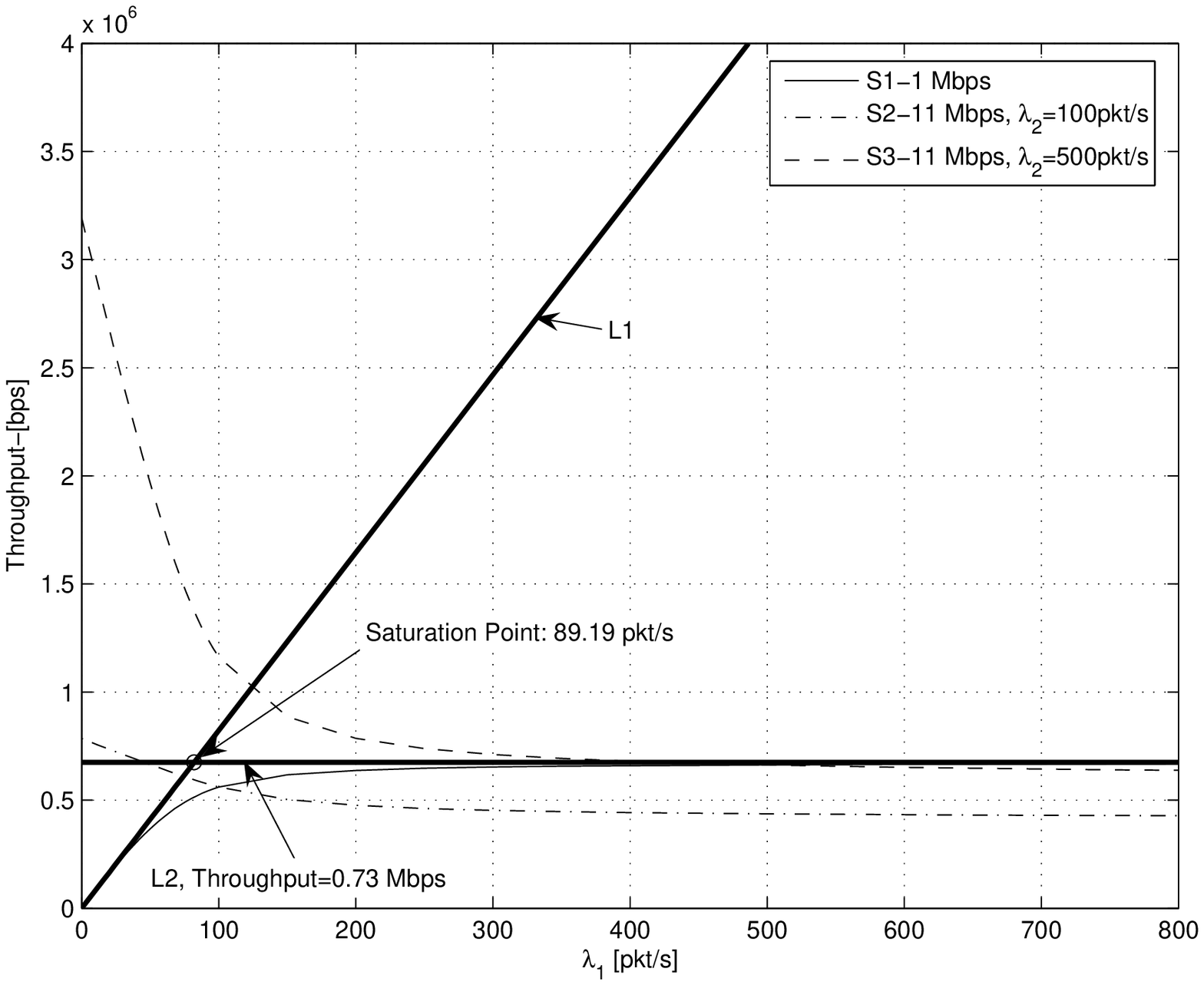}{Behaviour of the per-station
throughput in a network of three contending stations (labelled S1,
S2 and S3) as a function of the packet rate $\lambda_1$ of the
slowest station S1 transmitting at 1 Mbps. The other two stations
transmit at 11 Mbps with constant packet rates, respectively equal
to $100 \textrm{pkt/s}$ and $500 \textrm{pkt/s}$. The packet size
PL is equal to 1028 bytes for the three contending
stations.}{throughput_caso1}
\clearpage
\figura{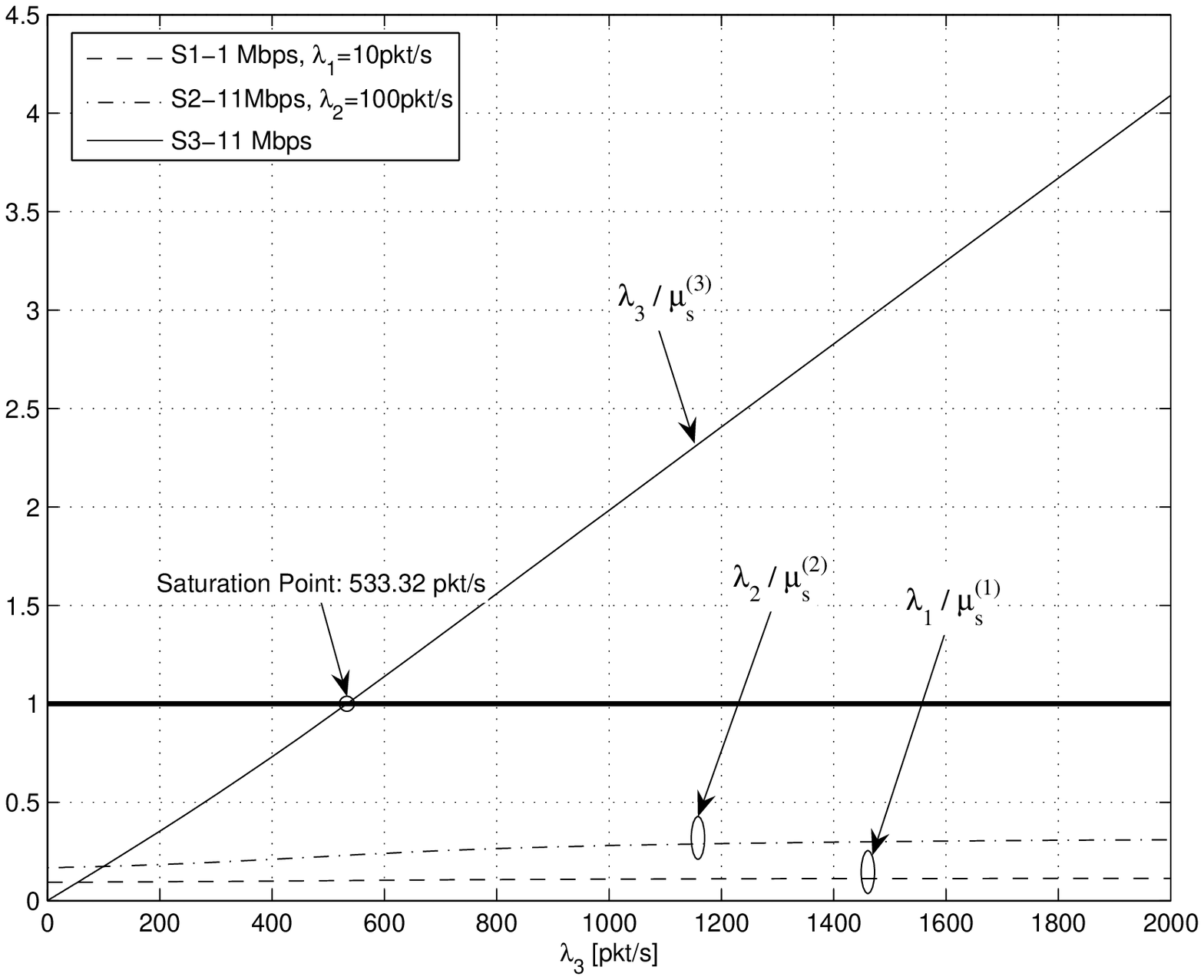}{Behaviour of the ratio
$\lambda_t/\mu_S^{(t)}=\lambda_t\cdot T^{(t)}_{serv}$ in
(\ref{saturation_condition}) in a network of three contending
stations (labelled S1, S2 and S3) as a function of the packet rate
$\lambda_3$ of the fastest station S3 transmitting at 11 Mbps. The
station S1 transmits with constant packet rate $10 \textrm{pkt/s}$
at 1 Mbps, whereas the station S2 transmits with constant packet
rate $100 \textrm{pkt/s}$ at 11 Mbps. The packet size PL is equal
to 1028 bytes for the three contending
stations.}{lambda_mu_s_caso3}
\clearpage
\figura{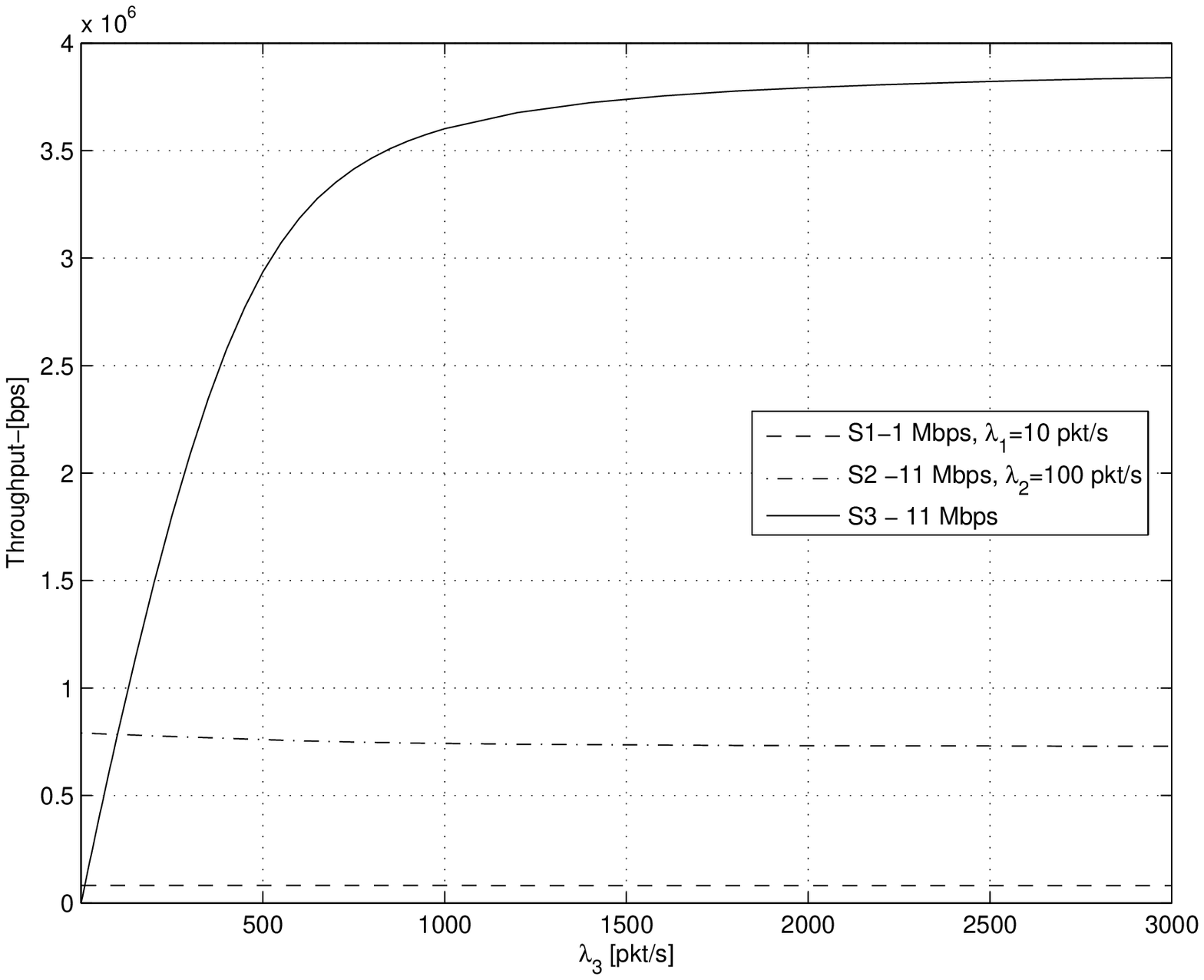}{Behaviour of the per-station
throughput in a network of three contending stations (labelled S1,
S2 and S3) as a function of the packet rate $\lambda_3$ of the
fastest station S3 transmitting at 11 Mbps. The other two stations
transmit with constant packet rates, respectively equal to $10
\textrm{pkt/s}$ and $100 \textrm{pkt/s}$. The packet size PL is
equal to 1028 bytes for the three contending
stations.}{throughput_caso3}
\clearpage
\begin{table}\caption{Typical network parameters}
\begin{center}
\begin{tabular}{c|c||c|c}\hline
\hline MAC header & 28 bytes&Propag. delay $\tau_p$ & 1 $\mu s$\\
\hline PLCP Preamble & 144 bit & PLCP Header & 48 bit \\
\hline PHY header & 24 bytes&Slot time & 20 $\mu s$\\
\hline PLCP rate & 1Mbps & W$_0$ & 32\\
\hline No. back-off stages, m & 5 & W$_{max}$ & 1024\\
\hline Payload size & 1028 bytes&SIFS & 10 $\mu s$\\
\hline ACK & 14 bytes&DIFS & 50 $\mu s$\\
\hline ACK timeout & 364$\mu s$ &EIFS & 364 $\mu s$\\

\hline\hline
\end{tabular}
 \label{tab.design.times}
\end{center}
\end{table}
\clearpage
\figura{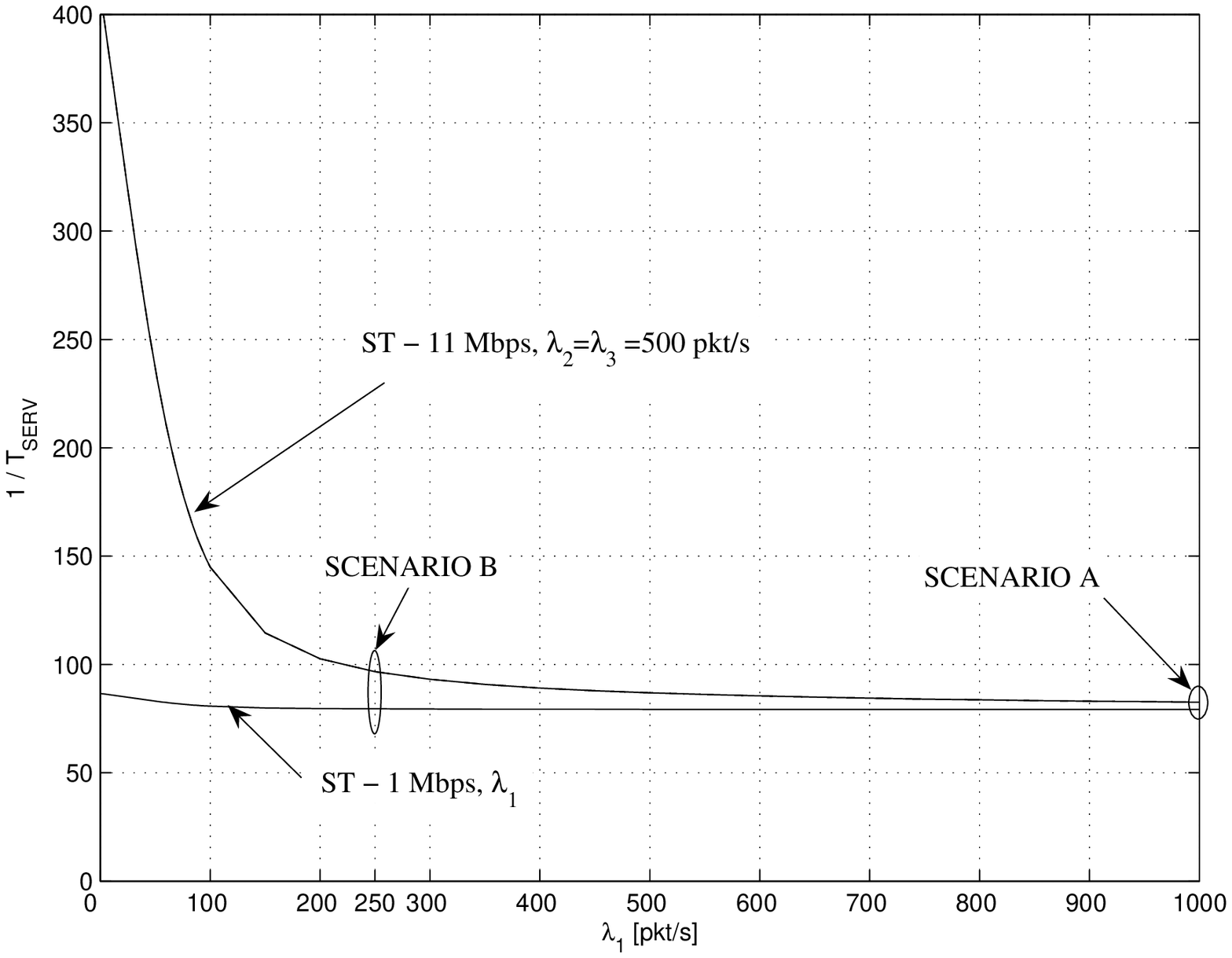}{Behaviour of the critical packet rates
($\mu_S^{(t)}=1/T^t_{serv}$) in a network of three contending
stations as a function of the packet rate $\lambda_1$ of the
slowest station transmitting at 1 Mbps. The other two stations
transmit with a constant packet rate equal to $500 \textrm{pkt/s}$
at 11 Mbps. Notice that curves $\mu_S^{(t)}$ related to the
stations at 11 Mbps are superimposed since they both employ the
same network parameters.}{behav_Tserv}
\clearpage
\figuramedia{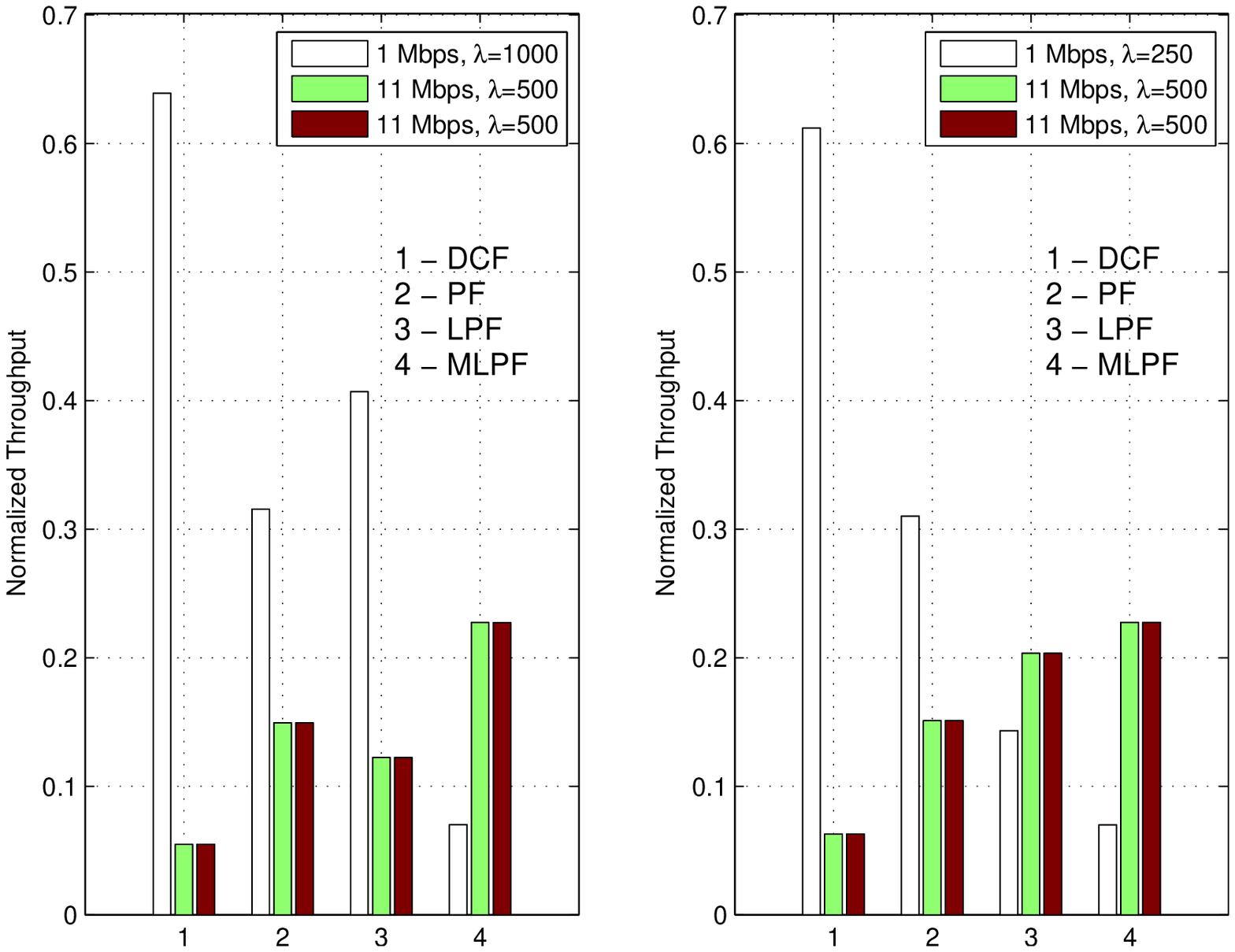}{Simulated normalized throughput
achieved by three contending stations upon employing 1) a
classical DCF; 2) DCF with PF allocation; 3) DCF optimized as
noted in (\ref{eq:max_prop_fair_lambda}); and 4) DCF optimized
with the MLPF criterion. Left and right plots refer to scenarios A
and B, respectively.}{fig_PF_1}
\clearpage
\figura{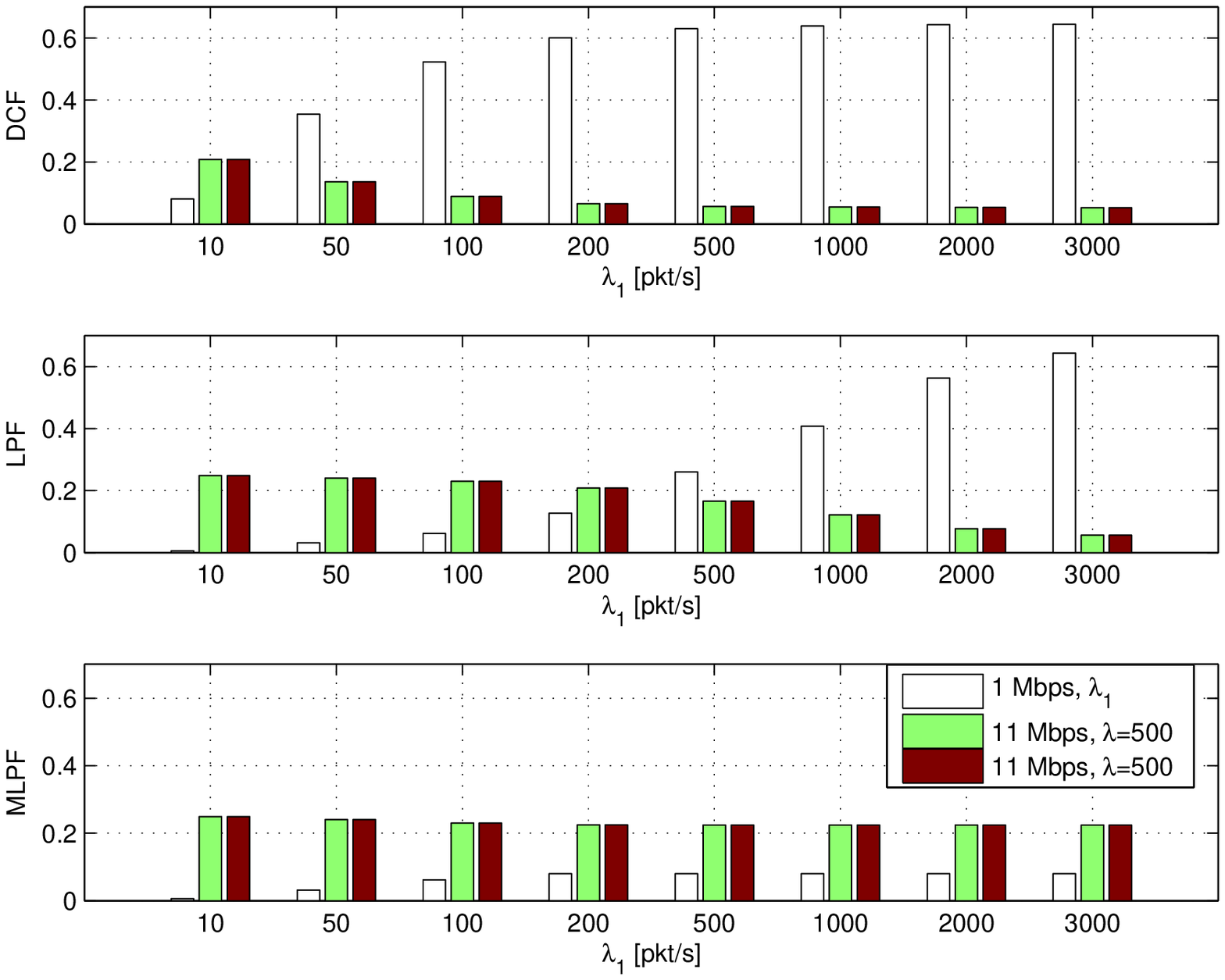}{Simulated normalized throughput
achieved by three contending stations as a function of the packet
rate of the slowest station in DCF, LPF and MLPF
modes.}{ThroughDCF_simfin}
\end{document}